%% file: paper.tex
\documentclass[sigplan,nonacm]{acmart}

\usepackage[table,dvipsnames]{xcolor}
\usepackage{cleveref}
\begin{document}

\title{Goldilocks Isolation: High Performance VMs with
Edera}
\renewcommand{\shorttitle}{Goldilocks Isolation}


 \author{Marina Moore}
 \affiliation{
   \institution{Edera}
   \country{}
}
 \email{marina@edera.dev}

 \author{Alex Zenla}
\affiliation{
	\institution{Edera}
	\country{}
}
	\email{alex@edera.dev}


\begin{abstract}
Organizations run applications on cloud infrastructure shared between multiple users and organizations. 
Popular tooling for this shared infrastructure, including Docker and Kubernetes, supports such multi-tenancy through the use of operating system virtualization.
With operating system virtualization (known as containerization), multiple applications share the same kernel, reducing the runtime overhead and allowing for many applications to run on the same machine.
However, this shared kernel presents a large attack surface and has led to a proliferation of \textit{container escape} attacks in which a kernel exploit lets an attacker escape the isolation of operating system virtualization to access other applications or the operating system itself.
To address this, some systems have proposed a return to hypervisor virtualization for stronger isolation between applications.
However, no existing system has achieved both the isolation of hypervisor virtualization and the performance and usability of operating system virtualization.

We present Edera, a VMM built for containers that uses paravirtualization, a minimal hypervisor, and dynamic resource management to enable modern cloud systems to use hypervisor virtualization without sacrificing performance.
We illustrate Edera's usability and performance 
by creating a container runtime compatible with Kubernetes that runs on the Edera hypervisor.
This implementation can be used as a drop-in replacement for the Kubernetes runtime and is compatible with all the tooling in the Kubernetes ecosystem. 
We find that Edera has runtime comparable to Docker with 2.14\% slower cpu speeds and faster memory performance.
It achieves this with a 740 millisecond increase in startup time from Docker's 381 milliseconds.
\end{abstract}




\maketitle

\input{introduction}
\input{background}
\input{requirements}
\input{design}

\input{implementation}

\input{analysis}
\input{discussion}

\input{conclusion}

\bibliographystyle{ACM-Reference-Format}
\bibliography{bibliography.bib}
\input{appendix-kata}

\end{document}

%% file: introduction.tex
\section{Introduction}
\label{section:introduction}

Cloud computing allows applications to share infrastructure, reducing the cost of deployment.
Using cloud resources, organizations can use hardware more efficiently, 
scale resource usage up and down with traffic, and provision their software quickly.
In the early days of cloud computing, applications were run on virtual machines~\cite{xen}, which gave each application its own operating system (OS) kernel.
Developers could run applications on this virtual machine, with a \textit{hypervisor} managing the hardware sharing between many virtual machines.
Now, it is more common to run multiple applications on the same kernel using operating system virtualization (OS virtualization) to separate the runtimes of each application~\cite{lxc, solaris-zones, apiary}.
This transition is due to both the efficiency of OS virtualization and the improved ease of use from a large ecosystem of tools available to use OS virtualization to run \textit{containers}.

Containers are an abstraction over applications that contain the application and its dependencies.
They are used by orchestration frameworks like Kubernetes to facilitate the use of OS virtualization.
The OS runs a container engine that manages the creation and operation of multiple containers.
This is a powerful abstraction that has been used to add efficiency, monitoring, and security to OS virtualization.
By 2027, Gartner predicts that 90\% of organizations will be running containerized applications in production~\cite{gartner-container-usage}.
Many of these organizations use OS virtualization in \textit{multi-tenant} environments, with containers from different users or companies running on the same host OS.
Although resource sharing enabled by OS virtualization has improved application runtime and resource usage, the shared host OS means it comes at the cost of strong isolation between applications.

OS virtualization uses kernel isolation techniques such as namespaces~\cite{namespaces}, cgroups~\cite{cgroups}, seccomp~\cite{seccomp}, and capabilities~\cite{capabilities} to  limit a container's access to kernel resources.
Over time, containers have added more of these isolation techniques, starting with just namespaces and adding layers to create more boundaries around the container runtime.
However, the nature of the shared kernel leaves a large attack surface for \textit{container escapes} in which an attacker running an application in a container is able to gain access to the kernel or other containers, ``escaping'' the isolation of the container runtime.
These attacks are frequent~\cite{firecracker, Unikernels:paper, lightvm, pvm, vkernel, CVE-2022-0847, CVE-2022-0492, CVE-2022-0185, CVE-2022-23648, CVE-2022-0811}.


In order to prevent container escapes, virtualized applications need stronger isolation, without sacrificing the runtime and ease-of-use present in modern virtualization systems.
We find that container escape vulnerabilities usually include a kernel exploit that an attacker uses to circumvent kernel isolation techniques.
These attacks can be prevented by moving the shared kernel out of the trusted computing base (TCB).
A return to hypervisor 
virtualization
provides a kernel for each application, providing stronger isolation.
By eliminating the shared kernel, an attacker who compromises the container will not gain access to the host machine or other applications without also compromising the hypervisor.
However, any solution that provides strong isolation will not be adopted unless it can match the runtime, resource usage, and ease-of-use of OS virtualization.

Previous work in container isolation has made progress toward achieving this combination of strong isolation and ease-of-use.
Unikernel approaches~\cite{Unikernels:paper, lightvm, unikraft, hermitux, mirage} combine the application and kernel into a single light-weight workload.
However this approach generally requires re-building applications for use in a unikernel, as well as on every iteration of the application, hindering development speed.
Other virtual machine approaches either require specialized hardware~\cite{kata, firecracker, pvm, oxide, metalvisor, constellation}, such as CPUs with virtualization extensions, or have high performance costs~\cite{gvisor}. 
While these approaches have been used for security-critical applications, these limitations have prevented their widespread adoption.

In this work, we observe three things about the current state of cloud computing.
First, containers have added many layers of segmentation to act more like a partition of computing resources, but OS virtualization does not have access to hardware partitioning.
Second, paravirtualization~\cite{Denali, xen} allows hypervisor virtualization on existing cloud infrastructure, without the need for hardware virtualization extensions.
Paravirtualization 
utilizes \textit{hypercalls} on guest OSs, replacing some system calls and avoiding costly emulation of I/O components.
Paravirtualization has been optimized for cloud use cases~\cite{cooperative-paravirtualization, paravirt-numa} making it ideal for improving the isolation of containers.
Third, recent upstream Linux support for paravirtualization enables existing kernels to run with hyprevisor awareness, without the need for full hardware emulation.

Based on these observations, we present the Edera VMM, an orchestrator for a type 1 hypervisor that uses a minimal design, memory safety, and dynamic resource management to provide strong isolation between containerized applications without sacrificing performance or usability.
Edera uses the Xen~\cite{xen} microkernel hypervisor to initialized a hardened root zone that controls the memory and CPU usage.
The use of Xen's type 1 hypervisor gives Edera hardware backed partitioning that is not supported by either containers or a type 2 hypervisor. 
This root hardened zone contains novel VMM tooling that utilizes disagregation and dynamic resource management.
Disagregation separates resources into isolated environments where possible, including drivers and system services.
Dynamic resource management utilizes insight from the type 1 hypervisor into running workloads to perform optimizations.
These optimizations include shared read-only kernel pages, matching the resource needs of containers, and preventing crashes due to usage spikes. 
Edera introduces \textit{zones} that encapsulate each isolated area of the system, including drivers and the guest operating systems.
These zones support minimal kernel images, such as those from the Open Container Initiative (OCI). 
While OCI images are not full micro kernels, they provide improved performance over full virtual machines.
Both the microkernel design and the use of Rust for memory safety reduce the hypervisor attack surface.

We test Edera's usability and flexibility by implementing it along with a container runtime compatible with the Kubernetes specification.
Kubernetes compatibility provides interoperability with the full range of compatible tools for scalability, monitoring, and more.
This implementation demonstrates that Edera provides the same ease-of-use as systems with OS virtualization and can be used as a drop-in replacement for popular container runtimes.


In summary, our contributions are:
\begin{itemize}
    \item We present Edera, an optimized VMM built for containers.
    \item We assess the isolation and performance of Edera, and find that it has 2.14\% slower CPU speed, faster memory performance, and 740 milliseconds of additional startup time compared to using Docker, while providing strong isolation. Compared to systems with strong isolation, Edera outperforms gVisor in and has comparable performance to Kata Containers and Firecracker with Edera performing better on CPU and memory operations and worse on system calls.
    \item 
    We implement a drop-in Kubernetes container runtime replacement that gives users the ease-of-use of modern OS virtualization systems. 
\end{itemize}

%% file: background.tex
\section{Background}
\label{section:background}

\begin{figure}
    \centering
    \includegraphics[width=.9\linewidth]{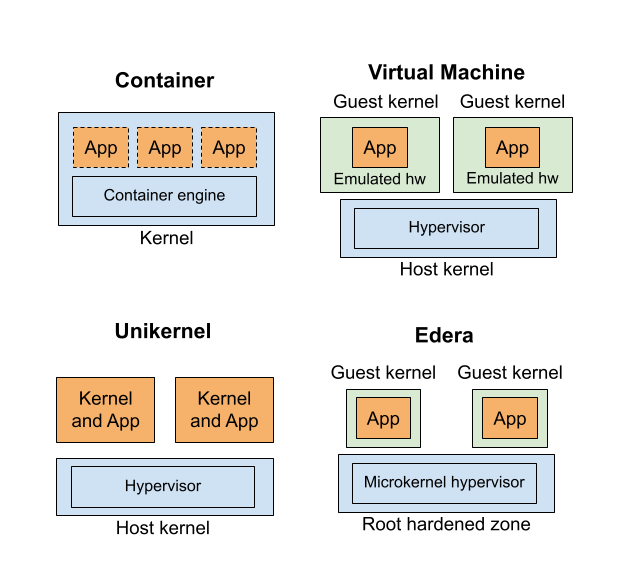}
    \caption{A summary comparison of virtualization designs. The containers in this diagram use OS virtualization. }
    \label{fig:isolation-overview}
\end{figure}

We provide background on the isolation and performance of virtualization techniques, summarized in \cref{fig:isolation-overview}.
We further provide background about Kubernetes to motivate our use case.

\subsection{Operating system virtualization}
OS virtualization
allows multiple containers to run on a shared kernel.
These containers include an application along with its runtime.
Examples of OS virtualization are Linux containers~\cite{lxc}, Solaris zones~\cite{solaris-zones}, and Apiary~\cite{apiary}.
Containers simplify deployment and scaling of applications by allowing multiple copies of an application to be dynamically added or removed, and a large ecosystem has been built around their use.
However, there is no strong isolation between containers running on the same infrastructure.

OS virtualization uses Linux isolation techniques to prevent containers from interfering with each other, including the use of \textit{namespaces}~\cite{namespaces}, \textit{cgroups}~\cite{cgroups}, \textit{seccomp}~\cite{seccomp}, \textit{capabilities}~\cite{capabilities}, and \textit{SELinux}~\cite{selinux}.
However, as containers still operate on the same kernel, many attacks have circumvented these isolation techniques to gain access to the kernel or other containers running on the same machine~\cite{firecracker, Unikernels:paper, lightvm, pvm, vkernel, CVE-2022-0847, CVE-2022-0492, CVE-2022-0185, CVE-2022-23648, CVE-2022-0811}. 
To prevent these \textit{container escape} attacks, systems require stronger isolation.

\subsection{Hypervisor virtualization}
Hypervisor virtualization uses a separate ``guest'' OS for each virtual machine (VM) with a hypervisor 
creating and monitoring VMs. 
This provides a separate kernel for each application, ensuring stronger isolation.
In a type 1 (or bare metal) hypervisor, such as Xen~\cite{xen} or VMware, the hypervisor runs directly on the host hardware.
Type 2 (or hosted) hypervisors such as VirtualBox instead run as a process on a host OS. 
Many hypervisors use x86 virtualization, emulating all hardware that the guest OS expects.
Paravirtualization~\cite{Denali, xen} instead requires the guest OS to make \textit{hypercalls} to the hypervisor for privileged operations.
This hypervisor-aware design avoids costly emulation of I/O hardware components.

\textbf{Isolation Comparison}
Unlike hypervisor virtualization, OS virtualization includes the shared kernel in the trusted computing base (TCB).
With this larger TCB, OS virtualization has weaker isolation.
This difference can be seen in practice in the abundance of container escape attacks on systems using OS virtualization~\cite{CVE-2022-0847, CVE-2022-0492, CVE-2022-0185, CVE-2022-23648, CVE-2022-0811, CVE-2024-0132}.
These attacks occur despite the use of Linux isolation techniques because the kernel is part of the TCB, and applications are thus vulnerable to flaws in the kernel design.
The shared OS and larger TCB increases the attack surface of OS virtualization, allowing privilege escalation, side channel attacks, and information leakage between containers.
Despite this, OS virtualization is common in practice because of its speed, cost savings, and the large ecosystem built around container orchestration technologies like Kubernetes.


\subsection{Additional Isolation Techniques}

\textbf{Hypervisor virtualization.} Several systems improve the efficiency of hypervisor virtualization.
Unikernels~\cite{Unikernels:paper, lightvm, unikraft, hermitux, mirage} use a minimized, per-application kernel to reduce the overhead of creating VMs.
This approach generally (with the exception of HermiTux~\cite{hermitux}) requires re-building images specifically for the unikernel, which can be slow and may be impossible for proprietary applications without source code access.
A unikernel also means that changes to images require re-compilation of the unikernel, hampering development speed.

Hardware virtualization extensions can improve hypervisor efficiency. Kernel-based Virtual Machine (KVM) is a Linux kernel feature built on hardware virtualization extensions that allows the kernel to function as a hypervisor~\cite{kvm}.
Firecracker~\cite{firecracker} and PVM~\cite{pvm} are virtual machine monitors (VMMs) that manage the device model and creation of virtual machines on the KVM hypervisor.
However, reliance on virtualization extensions in the underlying KVM means that these systems cannot run in clouds that do not support nested virtualization.
There are also performance impacts, especially for startup time that have hindered adoption in all but the most security-sensitive applications.
Further, KVM is built on the Linux kernel, so includes this whole kernel in the TCB. 
Operation forwarding attacks on Kata~\cite{kata} and Firecracker demonstrate vulnerabilities with the use of KVM to forward operations to the host kernel~\cite{operation-forwarding}. 

Other hypervisor virtualization techiques have been introduced for the cloud. Xen-blanket~\cite{xen-blanket} adds an intermediate compatibility layer to cloud virtual machines. This provides performant interoperability, but does not directly run containers.

\textbf{OS Virtualization.} Other isolation techniques add isolation layers to OS virtualization.
These include gVisor~\cite{gvisor} which implements user-level kernel isolation, and vKernel~\cite{vkernel} which allows for container-specific security rules.
However, these systems do not achieve the isolation of hypervisor virtualization as they still contain a shared kernel and filesystem, which attackers can use to discover sensitive data~\cite{gvisor-gemini-attack}.

\textbf{Dedicated hardware.} Other approaches such as Oxide computers~\cite{oxide} and Metalvisor~\cite{metalvisor} improve container isolation with dedicated hardware.
Constellation~\cite{constellation} and Confidential Containers~\cite{confidential-containers} use hardware trusted execution environments (TEEs) to isolate Kubernetes clusters.


\subsection{Kubernetes}
Kubernetes~\cite{kubernetes} is a system for automatically deploying, scaling, and managing containerized applications.
It runs containers in \textit{pods} that run on \textit{nodes}.
A \textit{container runtime} manages the lifecycle of pods.
Kubernetes can work with any container runtime that follows the Container Runtime Interface (CRI) specification~\cite{cri}.
One such container runtime is \textit{runc}, a lightweight runtime based on Docker.
A \textit{container runtime interface}, such as \textit{containerd} provides higher-level control of containers on the node, including image management and container execution. 
The Kubernetes API server lets users query and change the state of Kubernetes across multiple nodes.
A \textit{kubelet} is an agent that runs on each node that registers the node with the API server and ensures that all pods expected by the API server are running.

The Kubernetes ecosystem is composed of projects such as container runtimes, container image distribution services, and projects that add functionality or usability. 
The Cloud Native Computing Foundation alone hosts more than 200 open source projects in this ecosystem~\cite{cncf-project-metrics}.
The scope of this ecosystem means that any incompatible new technology will be at a disadvantage as it would lack compatibility with Kubernetes and related projects as well as pre-built OCI container images.

Systems have been built to allow Kubernetes to manage VMs instead of or in addition to OS virtualized containers.
This provides Kubernetes compatability to hypervisor virtualization.
Many of these, including Infranetes~\cite{infranetes}, rkt~\cite{rkt}, and virtlet~\cite{virtlet} 
require virtualization extensions.
RunX~\cite{runx} uses a series of shell scripts to run VMs on Xen, but does not make optimizations in Xen or the VMM to improve performance.
These projects demonstrate the desire for strongly isolated hypervisor virtualization systems compatible with Kubernetes, but none meet our performance and usability requirements.


%% file: requirements.tex
\section{Requirements}
\label{section:requirements}

\newcommand{\good}{{\textcolor{ForestGreen}{\checkmark}}}
\newcommand{\warn}{{\textcolor{BurntOrange}{\danger}}}
\newcommand{\bad}{{\textcolor{Maroon}{$\times$}}}
%
%

Our goal is to create a system with strong container isolation while ensuring compatibility with the container ecosystem and performance comparable to OS virtualization. To achieve this, our requirements are as follows.

\textbf{Eliminate the shared kernel.} The system must support secure multi-tenancy, with the container acting as a security boundary. This means removing the kernel from the TCB so that a bug in the kernel is not sufficient to escape the container's isolation.

\textbf{Use existing OCI images.} Existing OCI container images must run without modification. Some OCI images have proprietary code which cannot be accessed by users to re-build the image. Even when the source is available, re-building images takes time and compute resources

\textbf{Cloud compatibility.} The system must run on commercially available systems without relying on specialized hardware like virtualization extensions or TEEs. Virtualization extensions are not available on all commercial cloud infrastructure as most cloud instances are already virtual machines, so access to virtualization extensions is only possible if the instance includes nested virtualization that passes through access to CPU features or if the instance provides bare metal access to a machine. Prices for a bare metal EC2 instance start at more than \$4/hour, compared to < \$0.01 for the smallest non-bare metal instance.


\textbf{Memory Safety.} The system should be written in a memory safe language to reduce the attack surface. Memory safety vastly reduces the number of memory leaks and memory corruption attacks.

\textbf{Runtime performance.} The system's runtime performance including for workloads heavy in CPU, memory, or system call usage, must be similar to that in existing systems. This ensures that it will not require more resources or higher costs to run.  The overhead of virtualization has been measured in previous work \cite{huber2011evaluating, younge2011analysis} with 5\% being the lowest overhead reported, so we require that our performance be within 5\% of running runc on Docker.

\textbf{Startup performance.} The system's startup occurs just once per application, but speed is important when provisioning new containers and scaling the number of copies up and down.
The startup for our system includes the VMM boot, guest OS boot and container runtime to be comparable with the container startup in Docker. Given this some overhead will be expected, but we set the goal of startup performance within 1 second of Docker, improving on the application start time in other hypervisor systems.


\subsection{Threat model}
\label{section:threat model}

In multi-tenant environments, workloads from third parties are run on shared infrastructure. 
The infrastructure allows users to run their own code, but the operator does not want users to gain access to the underlying infrastructure or other workloads.
Given this environment, for our threat model we assume an attacker can:
\begin{itemize}
    \item Run an application in a zone on a multi-tenant machine.
    \item Have root access to the Linux kernel in that zone.
    \item Run any Linux system call in the zone.
\end{itemize}

\noindent In this environment, an attacker succeeds if they are able to:
\begin{itemize}
    \item Access another zone, the hypervisor, or root zone.
    \item View memory, CPU usage, or other information about containers running in other zones.
    \item Gain access to a running zone that they do not control.
\end{itemize}

\noindent The following are out of scope:
\begin{itemize}
    \item Supply chain attacks on the hypervisor or root zone.
    \item Hardware attacks that require physical access to the machine.
    \item Side channel attacks, for instance those that use the temperature of the host machine. These attacks are unavoidable on shared infrastructure. 
\end{itemize}

%% file: design.tex
\section{Design}
\label{section:design}

Edera uses a type-1 paravirtualized \textbf{microkernel hypervisor} that creates a \textbf{root hardened zone} that acts as a VMM, passing memory and CPU from the host machine to zones.
The root hardened zone runs workloads in \textbf{zones}, which are fully functional VMs that run standard Linux for container workloads. The root hardened zone monitors for anomalies and can terminate other zones. The VMM controls and optimizes the memory and CPU usage of zones. Each zone is scheduled by the hypervisor with parameters from the root hardened zone like a user-space process within the hypervisor.


\subsection{Hypervisor}
\label{section:hypervisor}
The Edera hypervisor is composed of Xen~\cite{xen} as the microkernel hypervisor, and our novel VMM.
We call these elements collectively the Edera hypervisor.
The Xen microkernel acts as the host kernel and can run directly on the host machine or as a guest OS in nested virtualization. 
It is a microkernel as the hypervisor only controls essential hardware interaction, while other components such as networking and device drivers are left for user-space processes in the root hardened zone and other isolated zones.
Research has shown that microkernels have fewer bugs, as many bugs come from less used parts of the kernel~\cite{popular-paths}.
The hypervisor is solely responsible for partitioning resources to each zone, and monitoring the usage of these resources.

Xen forms the base of Edera, but we utilize only the core components and not the orchestration or other tooling that are part of the Xen project.
Xen uses MISRA C and boasts a mature codebase that makes it suitable for safety domains including automotive and aerospace \cite{xenproject-embedded, xen-aerospace, xen-automotive}.
Edera uses paravirtualization (PV) mode, with the hypervisor managing the CPU, MMU hardware page table, and network.
There is also support for PV with hardware virtualization (PVH) mode, which takes advantage of virtualization hardware for performance improvements. As PVH mode requires virtualization hardware, examples in this paper use PV mode unless otherwise stated.
Our build of Xen adds a small layer to simplify starting a new Xen domain, perform device enablement, and provide a Rust interface.

\subsection{Root hardened zone}
Edera's novel VMM, which we call the root hardened zone, provides a system for VM orchestration, device management, dynamic resource allocation, zone initialization, and inter-process communication (IPC).
The root hardened zone initializes the operating system in a zone, attaches disk and network devices to a zone, and performs RPC. 

\subsubsection{Resource manager}
The Edera resource manager monitors resource usage, detects misbehavior, and performs dynamic resource allocation.
The resource manager can also dynamically pin and unpin CPUs for each zone, and adjust the memory allocation.
This dynamic configuration allows for more efficient resource allocation as zones can be assigned CPUs and memory pages as needed, rather than these resources remaining static after zone creation.
This solves a common problem in containers, which need to be re-created if they run out of memory or CPU resources under static allocation.

Edera implements dynamic resource allocation that, to our knowledge, is the most expansive in any system.
While it is inspired by memory ballooning \cite{memory-ballooning} that can add or remove memory based on usage, we take this a step further by providing prioritization for each zone.
QubesOS performs similar prioritization, but its design assumes only one active VM at a time, and moves resources to this active VM.
Edera provides resource limits and prioritization for memory pages and CPU cores for multiple concurrently running zones.
These limits and priorities are initially set automatically, but can be updated by administrators as needed. 
Dynamic resource allocation takes advantage of Xen scheduling features to support non-standard memory designs such as non-uniform memory access (NUMA) \cite{xen-numa} to ensure efficient allocation even in environments with multiple CPUs.

\label{section:IPC}
\subsubsection{Inter-process communication}
IPC in Edera is handled with shared memory pages. 
When two zones need to communicate, for example when the root hardened zone needs to communicate with a newly created zone, the root hardened zone directs the hypervisor to allocate a page of memory and set it to be writable by both zones.
The hypervisor can thus ensure that only these zones can access this page using the grant table, which is ceded to the hypervisor from the root hardened zone on initialization.
To ensure the security of this communication, grant references can only be used by the zone they are given to.
This process ensures that no zones, including the root hardened zone, can read communication between zones set up through IPC.
Using the same process, driver zones attach to a zone by using the grant table to get shared pages with the zone.
The device can then communicate to the attached zone, and no other zones, using this shared page.

\subsection{Zones}
Each zone launched by the hypervisor runs a guest OS kernel.
Using PV, the hypervisor runs in ring 0, and all guest OSs run in a less privileged ring (ring 1 for x86 and ring 3 on x86-64)~\cite{xen-64}.
As most OSs expect to run in ring 0, the guest OS must be aware that it is running in a virtualized environment and support hypercalls.
Most modern Linux distribution meet these requirements.
These hypercalls~\cite{xen} are used for privileged operations that need to run in ring 0 and pass along those privileged system calls to the hypervisor.
This is supported by most OSs through a PV mode.

To optimize memory usage, guest kernels in Edera are placed in read-only memory that can be shared between multiple zones using the same operating system. These read-only kernel pages save system memory and prevent userspace programs from tampering with kernel functionality.

Edera supports running OCI images as the guest OS.
Many OCI images are built as a minimal operating system interface for containers, and so provide a smaller footprint than installing a full Linux kernel.
While not a true microkernel, these minimal images help improve the overhead of guest OSs in Edera.

\subsubsection{Inter-domain messaging}
Edera uses inter-domain messaging (IDM) to send commands or processes to the zone (such as through the Kubernetes utility kubectl exec), in addition to sending logs and metadata back to the hypervisor.
IDM uses a Xen byte channel~\cite{xen-channel} to send protobuf messages to other zones using IPC. 
Protobuf (shortened from protocol buffers) is a platform-neutral mechanism for serializing structured data.
Its design includes protection against overloading the pipe by using bound checks and other limits to ensure safe operation. 

\subsubsection{Monitoring}
An \textit{init} process in every zone is started when the zone is provisioned. This process runs in the background to monitor the VM.
It sends all monitored events over the IDM channel to the root hardened zone to be logged and collected.
If the init process is disabled, the root zone marks the zone as not responding and deprovisions it. 

Edera's monitoring happens outside of the zone and uses information from init, CPU and memory usage, and 
network traffic passed through the hypervisor.
As guest kernels do not have access to ring 0 without going through the hypervisor, all privileged operations can be limited and monitored.
All network packets leaving the VM are monitored when they are passed through to the hypervisor, so a VM cannot send malicious traffic that bypasses the monitoring.
This monitoring can be used to manage resource usage as well as to perform detection and response for processes running on the VM.
For example, a misbehaving zone could be quarantined and monitored to understand malicious activity.

%% file: implementation.tex
\section{Implementation}
\label{section:implementation}

\begin{figure*}[ht]
    \centering
    \includegraphics[width=\linewidth]{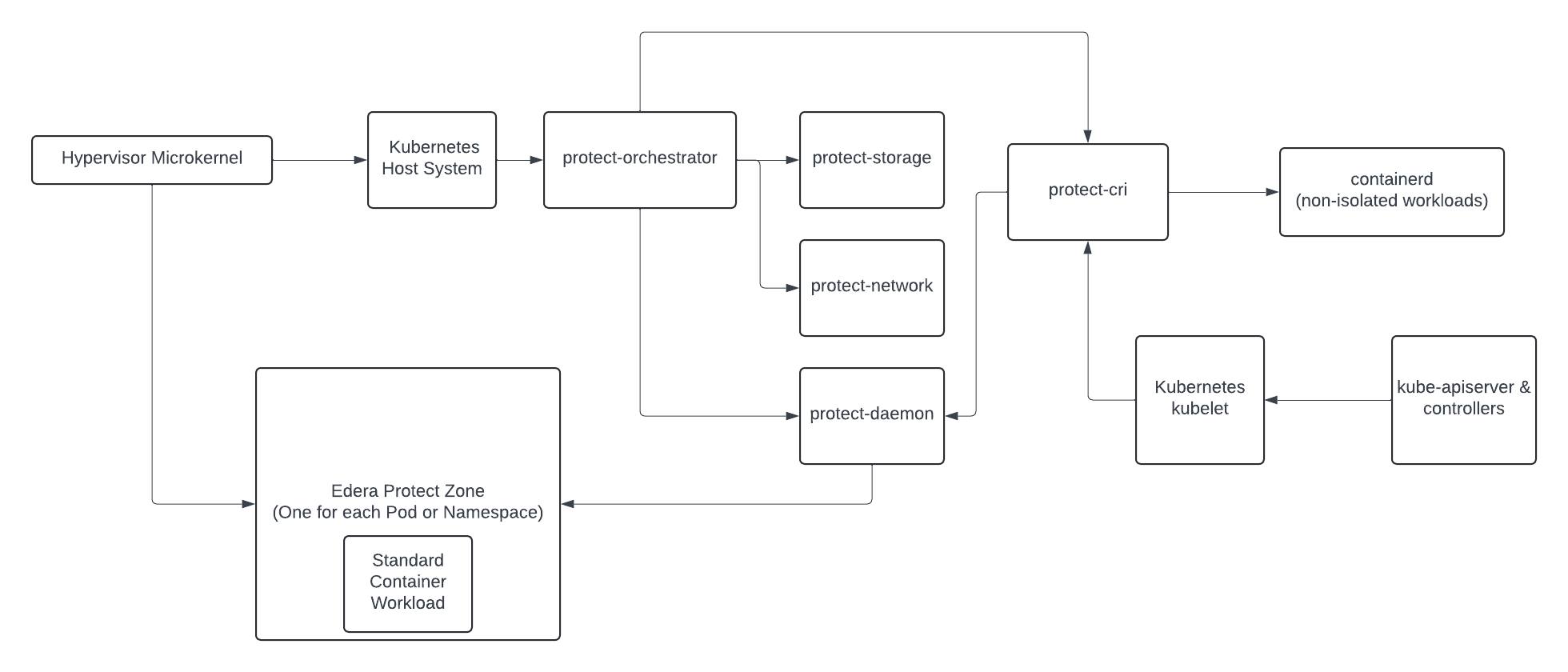}
    \caption{An overview of the services used by Edera to run a Kubernetes container.}
    \label{fig:services}
\end{figure*}

We implement Edera in Rust using the design from \cref{section:design} with built-in services for Kubernetes compatibility. 
By implementing it in Rust and utilizing MISRA C components from Xen, we avoid many memory safety bugs while achieving high performance.
The implementation includes our build of the Xen hypervisor, the root hardened zone, and several services that allow use with Kubernetes for cluster orchestration.
These services include a CRI compatible with the OCI runtime specification, daemons for zone monitoring, and a kube-apiserver for managing pods.
For ease of use with Kubernetes, we create a custom guest OS kernel that is distributed as an OCI image.
Thus, a user can specify this image in their Kubernetes configuration and automatically run their existing containers.
The source for this implementation is available on request.

We will walk through the system startup and running of a single Kubernetes pod in an Edera zone to illustrate the services used by the implementation, shown in \cref{fig:services}.
First, the machine boots into the microkernel hypervisor.
Xen microkernel starts the root hardened zone, which also acts as a Kubernetes host and runs protect-daemon.
The hypervisor launches protect-orchestrator, a management component that ensures that all other daemons (protect-network, protect-daemon, and protect-cri) are running.

\textbf{protect-daemon} implements the root hardened zone, keeping track of which zones are running each workload and managing drivers for the zones, while leaving resource management to the hypervisor.
This is the daemon that creates, modifies, and removes zones.
It manages an embedded key value store that tracks what is running in each zone where the zone is identified by a uuid that is assigned on zone creation.
In addition, protect-daemon manages the drivers for each zone, including IDM.
Using IDM, protect-daemon can pass commands from the API server (though protect-cri) to the init process in the zone.
It can also receive messages from the zone, and pass these along to protect-cri.



\textbf{protect-orchestrator} performs the initial system setup, and maintains running processes.
It provides updates and health checks for running daemons including protect-daemon, protect-network, and protect-cri.
In addition, it provides observability for all services running in the hypervisor, through tools like Prometheus~\cite{prometheus}.

\textbf{protect-network} performs packet routing and monitors network activity.
Zones have a basic ethernet interface, but cannot perform packet routing without going through protect-network.
protect-network receives tasks from zones through protect-daemon, then performs them on behalf of the zone.
It writes packets to the networking layer and can dial sockets for zones.
Zones do not have direct network access, so protect-network sees all routed packets, ensuring that Edera can monitor all network activity.

\textbf{protect-cri} provides a layer on top of 
any OCI-compliant CRI, and converts requests from the CRI to requests for protect-daemon.
protect-cri reconciles the state between protect-daemon and the CRI to ensure that the zones running on protect-daemon match the Kubernetes pods expected by the CRI.
It also transfers messages from zones through protect-daemon to other Kubernetes processes.
protect-cri supports running some Kubernetes nodes in Edera zones and others on different infrastructure by coordinating with containerd to run the non-Edera nodes.

Once the system is booted, a user may launch a pod from protect-cri.
This triggers the kubelet~\cite{kubelet}, which registers the pod with the kube-apiserver and launches a zone.

\textbf{Summary.} In summary, Edera provides services to manage zones and communicate with Kubernetes.
Each Kubernetes pod runs in its own Edera zone, ensuring no shared kernel state between pods.
By routing all network traffic through protect-network and all IDM communication through protect-daemon, we ensure that zones can be monitored. 


%% file: analysis.tex
\section{Analysis}
\label{section:analysis}

In order to analyze Edera, we compare it to other container isolation techniques in terms of performance and security.

\subsection{Performance}
\label{section:performance}

We compare the performance of Edera to existing systems for running containers. Docker using runc acts as a baseline metric for the performance of OS virtualization. gVisor, Kata Containers (Kata), and Firecracker allow comparison with other systems with strong isolation. We run gVisor with the default configuration, which uses seccomp-based systrap platform. We run Kata Containers using the KVM hypervisor with the Dragonball VMM. We choose Dragonball as it is optimized for container workloads \cite{kata-hypervisors}. See Appendix~\ref{appendix:kata} for some performance numbers that verify this.
We run Firecracker on the KVM hypervisor with the firecracker-containerd container runtime.
We especially compare Edera in PV mode to Docker and gVisor, as none of these systems require virtualization extensions, while comparing Edera PVH with Kata and Firecracker to compare performance when such extensions are available.
We run all experiments on a bare metal OVH machine with virtualization extensions.
The machine has 16 GB of memory and an Intel Xeon CPU with 8 cores at 3.20GHz.
There were 3 boot configurations we used to run these benchmarks: kvm, edera-pv, and edera-pvh. 
We ran benchmarks on all systems where they would run. Docker and gVisor run in userspace and are not hypervisor or operating system dependant, so these benchmarks ran in all three boot configurations.
Edera, Firecracker, and Kata depend on the hypervisor, so were only run on the appropriate systems.


We evaluate both startup and runtime performance.
Within runtime performance, we look at the CPU and memory usage, as well as the speed of various system calls.
To estimate real-world performance, we also evaluate the overall runtime for some example workloads.
All benchmarks are run on OCI container images (generally the latest Alpine image), with a fresh container created for each run to model the performance in a real-world cloud application.
Benchmarks are run with access to 4 vCPUs and 4096 MB of memory, except for the kcbench and pgbench which required more resources to run successfully on all systems.
We run benchmarks 5 times and present the average of these runs, with the standard error shown in all graphs.

\begin{table}[t]
	\centering
	\begin{tabular}{l c}
		\hline
		\textbf{system} & \textbf{startup time (ms)} \\
		\hline
		Docker (pv) & 381 \\
		Docker (pvh) & 378.8 \\
		gVisor (pvh) & 444\\
		gVisor (pv) & 452.2\\
		Docker (kvm) & 571.6 \\
		gVisor (kvm) & 684.4\\
		Edera-PV & 1120.8\\
		Kata Containers & 1913\\
		Firecracker & 2421.8\\
		Edera-PVH & 2516.8\\
		\hline
	\end{tabular}
	\caption{The startup time for each system. Rows are sorted by time.}
	\label{tab:startup}
\end{table}

\begin{figure*}[ht]
    \centering
    \includegraphics[width=\textwidth]{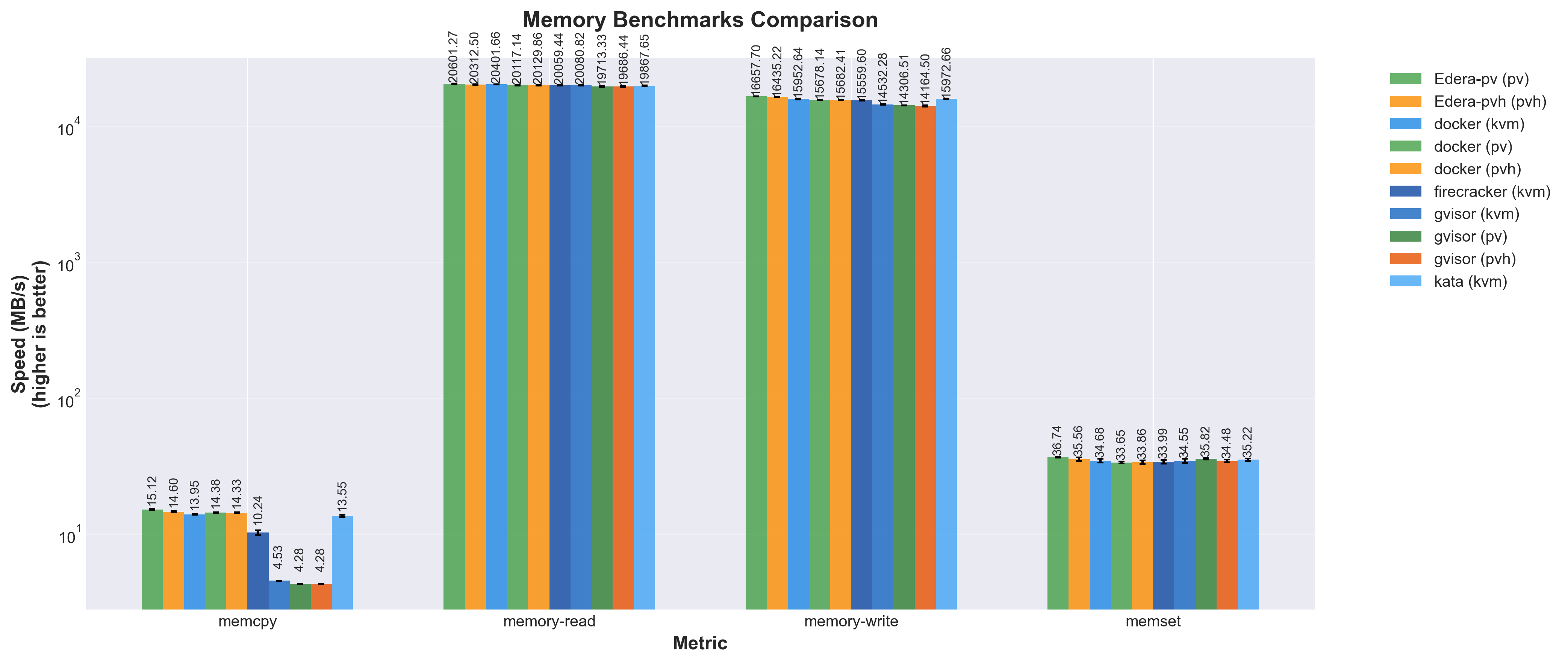}
    \caption{A comparison of the memory speed in GB/s using metrics from both sysbench and perf-bench. Higher is better.}
    \label{fig:memory}
\end{figure*}


\begin{table}[t]
	\centering
	\begin{tabular}{l c}
		\hline
		\textbf{system} & \textbf{Speed (events/sec)} \\
		\hline
		Firecracker & 183.8\\
		Docker (pvh) & 276.62 \\
		gVisor (pvh) & 310.84\\
		gVisor (pv) & 328.11\\
		Docker (pv) & 329.03 \\
		Edera-PVH & 359.41\\
		Edera-PV & 359.95\\
		Kata Containers & 366.67\\
		gVisor (kvm) & 366.89\\
		Docker (kvm) & 367.84 \\
		\hline
	\end{tabular}
	\caption{A comparison of the CPU speed in events per second using metrics from sysbench. Higher is better, results are sorted.}
	\label{tab:cpu}
\end{table}

\textbf{Startup time.} We first examine the container startup time, including the full time from VM init to starting the container, shown in \cref{tab:startup}. Container images were cached, so this does not include downlaod time. Edera takes 1.1 seconds to startup in PV mode, compared to 2.5 seconds in PVH mode. This is a slower startup time than Docker and gVisor on Linux, which start up in 0.57 and 0.68 seconds respectively, but PV is faster than Kata Containers and Firecracker, which start up in 1.9 and 2.4 seconds respectively. Note that this is slower than the Firecracker startup time reported in~\cite{firecracker} due to our reporting the complete container startup time rather than just the VM boot time. 

However, Edera zones, as well as Kata or Firecracker VMs, can be started before they are needed and kept running in an idle state until use. In this way, the user does not need to wait for the 1 second startup time to use a zone. With Edera's dynamic resource allocation these idle \textit{warm zones} can initially be allocated no memory or CPU resources, then resources can be dynamically allocated once the zone becomes active. 

\textbf{Memory and CPU benchmarks.}
We assess the memory and CPU speed of each system. Each system was configured with the same resources, so this helps compare their efficiency.
\Cref{fig:memory} shows the results from sysbench, a benchmarking tool that performs reads and writes to a data buffer, and the speed of memset and memcpy from perf-bench. 
sysbench used a block size of 1 GB and total memory size of 20 GB for memory operations and sequential write.
For the perf-bench memory benchmarks, the size was set to 100MB.

Edera outperforms the other systems on both sysbench tests, although by a small margin. 
Edera also demonstrates the best performance for memcpy and memset, with memcpy showing noticably slower results for Firecracker and gVisor. 
Docker performs similarly on bare metal and on the Edera hypervisor indicating minimal performance impact from the hypervisor on memory operations. gVisor performs slightly worse on the Edera hypervisor due to the additional indirection. In summary, all the systems, with the exception of gVisor and Firecracker have very similar memory performance.

\Cref{tab:cpu} illustrates the performance of the sysbench CPU benchmark.
For this test sysbench monitors CPU usage while checking prime numbers up to 150,000 over 4 threads. 
Edera-pv is 2.14\% slower than Docker on Linux.
Docker is 10-25\% slower on the Edera hypervisor, illustrating some performance loss through the use of vCPUs in Edera. Edera itself was faster on the hypervisor, illustrating that some of this performance is made up through the CPU management by Edera zones.
Firecracker performed at roughly half the speed, likely due to inefficient use of the configured 4 vCPUs.
More fine-grained CPU configuration to improve this is not supported in firecracker-containerd.

\begin{figure*}[t]
    \centering
\includegraphics[width=\textwidth]{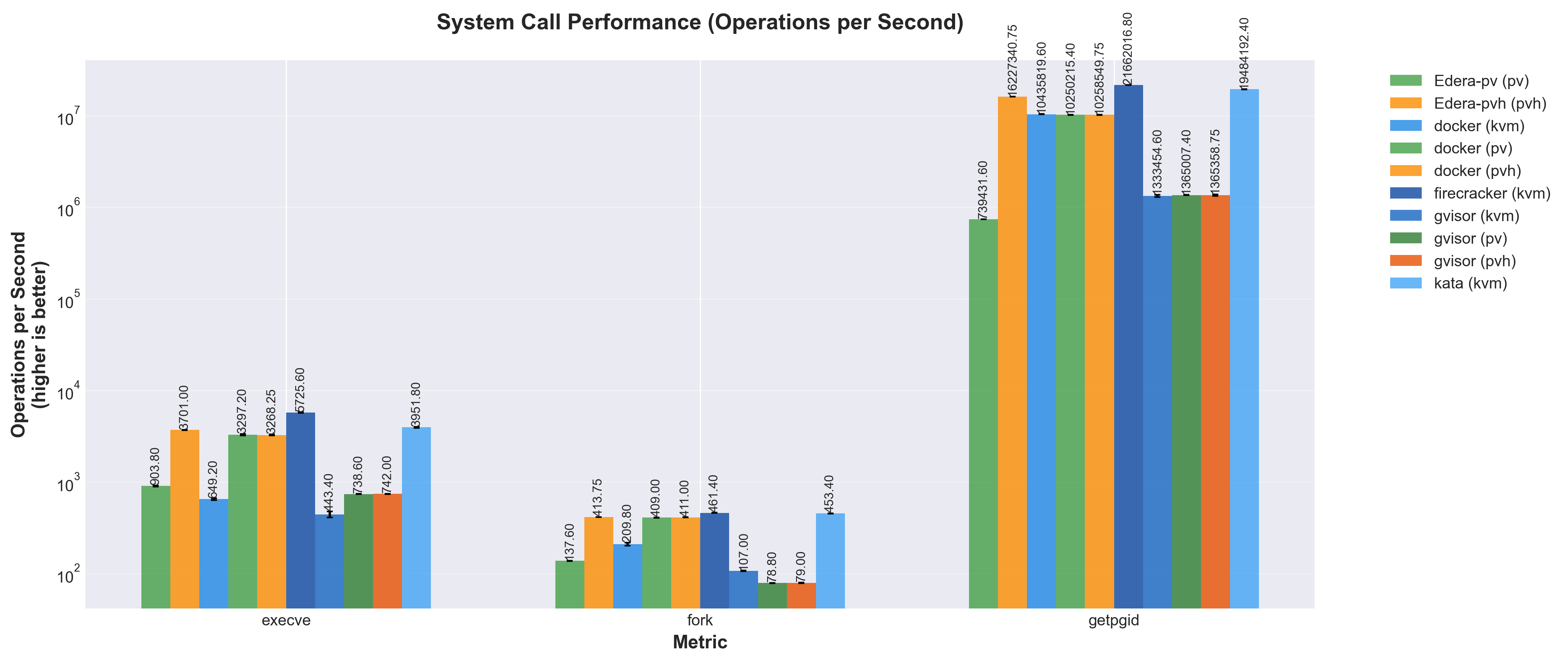}
    \caption{A comparison of the operations per second of system calls shown on a logarithmic scale. Higher is better.}
    \label{fig:syscall}
\end{figure*}

\begin{figure*}[t]
    \centering
\includegraphics[width=\textwidth]{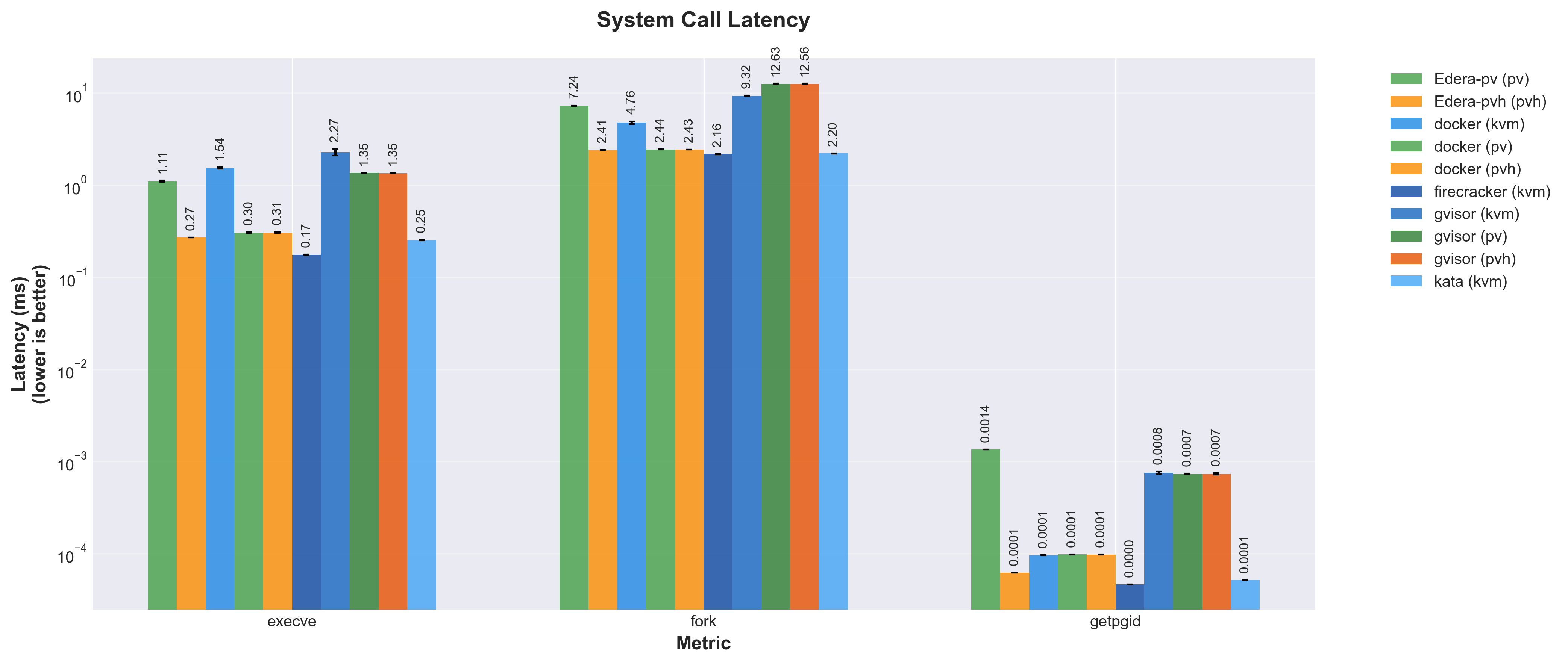}
    \caption{A comparison of the system call latency. Note that this graph is shown on a logarithmic scale. Lower is better.}
    \label{fig:syscall-latency}
\end{figure*}


\begin{figure}[t]
    \centering
    \includegraphics[width=\linewidth]{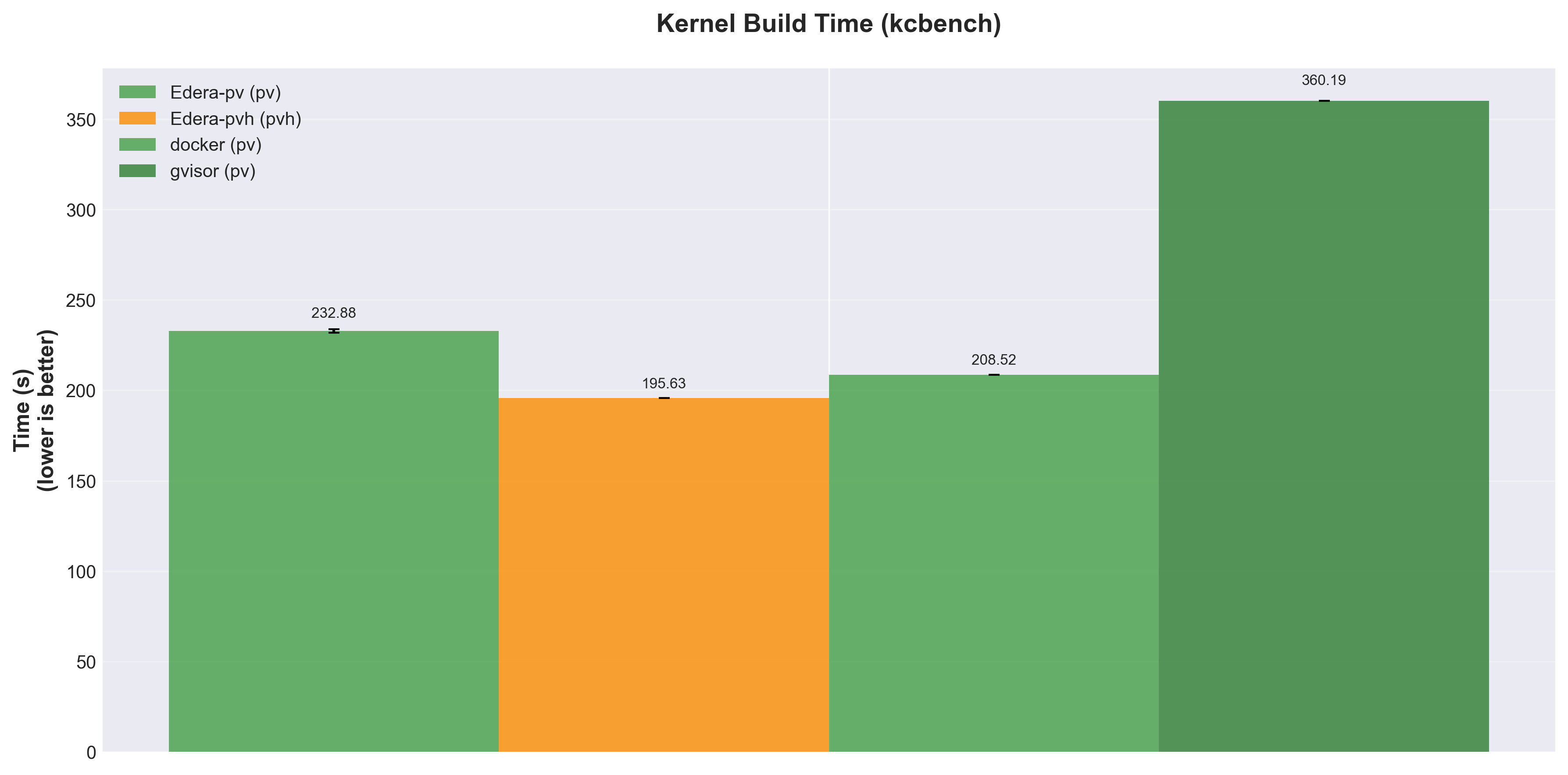}
    \caption{A comparison of runtime of kcbench in milliseconds. Lower is better.}
    \label{fig:kcbench}
\end{figure}

\begin{figure}[t]
    \centering
    \includegraphics[width=\linewidth]{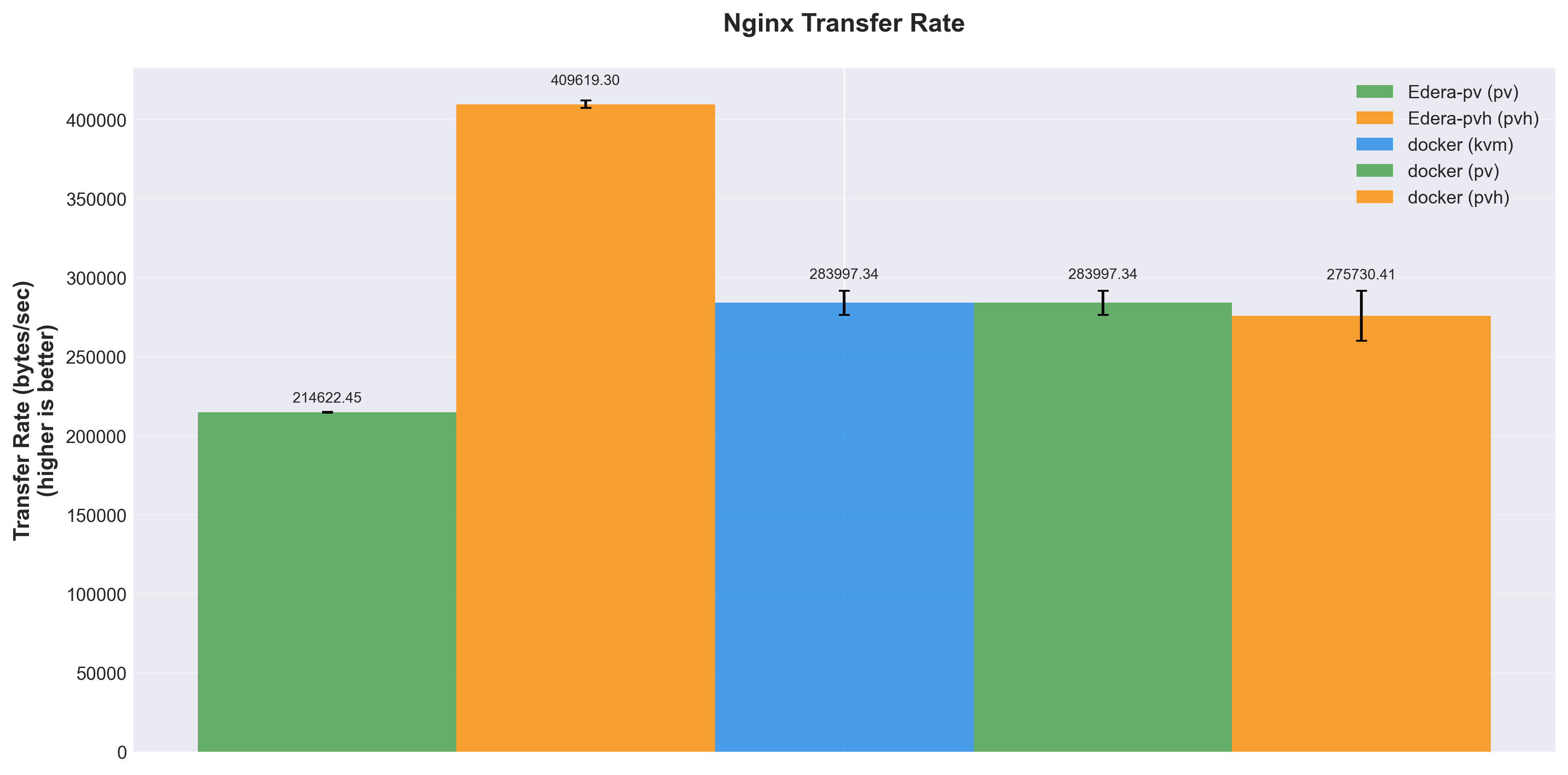}
    \caption{A logarithmic scale comparison of runtime of nginx in milliseconds. Lower is better.}
    \label{fig:nginx}
\end{figure}

\textbf{System calls.} Next, we evaluate the speed of executing several popular system calls.
The mechanism by which system calls are passed to the host system differs widely between the systems we evaluate, so we determine how these approaches impact performance.
\Cref{fig:syscall} shows the operations per second of each system running fork(), execve(), and getpgid() based on results from perf bench.
This graph shows that Edera outperforms gVisor in all system calls, and performs slower than Docker on bare metal (for example fork() is 41.8\% slower on Edera than Docker).
Edera avoids much of the indirection used by gVisor, leading to much faster performance.
However, with virtualization extensions, Edera PVH is able to outperform Docker (with fork() 169.4\% faster than Docker). 
This is slower than the system call performance of Kata and Firecracker, which also use virtualization extensions, with Kata 34.4\% faster and Firecracker 43.8\% faster than Edera PVH.

\Cref{fig:syscall-latency} further shows the latency of fork(), execve(), and getpgid().
Similar to the system call performance, the latency of Edera PVH is better than Docker and gVisor, but worse than Kata and Firecracker. 


\textbf{Real-world workloads.} 
Finally, we evaluate the runtime of workloads in each system to understand how they perform in real-world environments.
These runtimes are shown in \cref{fig:kcbench} and \cref{fig:nginx}.
We first evaluate kernel build time using kcbench.
We set the kernel version to 6.11.3 and 8 jobs in parallel. For this benchmark, all systems had access to 8 vCPUs and 16384 MB of memory.
Edera's kernel build time is 5\% slower than Docker, 59\% faster than gVisor, and 52\% faster than Kata.
A kernel build represents a more realistic workload with several different system calls and operations.
As this benchmark requires high memory usage, it does not run on firecracker-containerd.
Next, we evaluate the runtime of a simple nginx server using Apache Bench over 10,000 requests. 
We find that Edera outperforms gVisor by a factor of 10 (18 ms compared to 220 ms for gVisor), and performs comparably to Docker (15 ms).
Edera-PVH has slightly faster performance than Kata according to this metric, although Edera-PV is slightly slower.
The use of a webserver like nginx is a common application for cloud applications.

\subsubsection{Summary}
In summary, Edera outperforms gVisor in all performance metrics with the exception of startup time, while overall performing only slightly worse than Docker and similarly to Kata and Firecracker.
This means that Edera achieves strong application isolation with minimal impact on performance even for systems that do not have access to virtualization extensions.
The system startup time performs the worst, with 1.1 seconds of startup time for Edera compared to 0.38 seconds for Docker. Overall memory and CPU performance for Edera is similar to that of Docker, and system calls perform worse. 
When virtualization extensions are available, Edera in PVH mode has improvements to system call runtime over PV mode, and outperforms Kata and Firecracker on memory and CPU operations.

\subsection{Requirements}

\begin{table}[t]
    \centering
    \begin{tabular}{l c c}
         \hline
         \textbf{Attack} & \textbf{Protected?} \\
         \hline
         CVE-2022-0185 & \good \\
         CVE-2022-0492 &  \good \\
         CVE-2022-0811 (cr8escape) & \good \\
         CVE-2022-0847 (Dirty Pipe) & \good \\
         CVE-2022-23648 & \good \\
         CVE-2024-0132 &  \good \\ 
         CVE-2024-21626 (Leaky Vessels) & \good \\
         \hline
    \end{tabular}
    \caption{Overview of recent container escape CVEs and how they are mitigated by Edera.}
    \label{tab:attacks}
\end{table}

Next, we evaluate Edera using our requirements from \cref{section:requirements}.

\textbf{Eliminate shared kernel.} Edera uses a type-1 hypervisor to use CPU partitioning to isolate containers in zones that have their own guest OS. This separation eliminates the shared kernel used by OS virtualization.
 
 \textbf{Use existing OCI images.} Edera is compatible with OCI images, and protect-cri complies with the Kubernetes CRI specification. This ensures that existing applications, even those using proprietary software, can be run in Edera without modification.

\textbf{Cloud compatibility.} Edera does not rely on direct access to virtualization extensions. It uses paravirtualization on guest kernels to allow full use of Linux system calls without the need for hardware virtualization. 

\textbf{Memory safety.} Edera is written in Rust and uses MISRA C components from Xen, providing memory safety.

\textbf{Runtime performance.} Edera performs within 5\% of Docker by CPU and memory benchmarks. Some system calls were above this threshold, but when run in PVH mode Edera is faster than Docker by these metrics as well. 

\textbf{Startup performance.} Edera adds 0.74 seconds of startup time compared to Docker, meeting the requirement of less than 1 second of additional overhead. Edera's dynamic resource allocation means it can be further reduced with warm zones that are given resources when needed. 

\subsection{Security}

Next, we evaluate Edera's security by examining if it can prevent container escapes. 
We evaluate seven container escapes from 2022-2024 to determine how Edera fares against these attacks.
We focus on vulnerabilities in Linux, as these could impact both Edera and other isolation techniques and omit analysis of vulnerabilities specific to other isolation techniques.
\Cref{tab:attacks} summarizes the results.

\textbf{Findings.} We find that Edera prevents all attacks through the use of hypervisor virtualization.
Without a shared kernel, attacks on capabilities, namespaces, and other OS isolation techniques do not give an attacker access to the host machine. 
This means that any kernel exploit that causes a container escape will only impact the guest OS on Edera.
As such, Edera is not vulnerable to any of these past container escape attacks and is further not vulnerable to this class of attack in the future. We provide in-depth explanations of three of these CVEs to illustrate how Edera prevents them.

\textbf{CVE-2022-0492.} This vulnerability allows a user of cgroup release\_agent to bypass cgroup's namespace isolation, thus escalating privileges and causing a container escape. Edera does not use cgroup for isolation, instead placing the entire guest kernel in an Edera zone. As such, this vulnerability in cgroup cannot be used to escape an Edera zone.

\textbf{CVE-2022-0847.} This vulnerability, known as ``Dirty Pipe,'' was caused by the improper initialization of pipes.
Pipes could contain stale values, allowing an unprivileged user to write to read-only pages in the page cache and escalate their privileges. 
Edera provides guest OSs with read-only access to virtual page tables, which the hypervisor then maps to MMU hardware pages.
As each zone only has access to its own virtual page table, it a user in a zone is unable to write to pages in the hypervisor or other zones.

\textbf{CVE-2024-21626.} A vulnerability in runc, known as ``Leaky Vessels,'' uses a leak of an internal file descriptor to create a container with a working directory in the host filesystem namespace. This can be used to gain access to the host filesystem or create a privilege escalation by re-writing files on the host filesystem.
In Edera, guest OSs do not share a filesystem with the hypervisor, so even if a guest OS had access to a hypervisor file descriptor, they would not be able to access or overwrite files on the hypervisor.

\textbf{Summary.} These vulnerabilities illustrate the security gains of hypervisor isolation. By eliminating the shared kernel, including shared access to pages and the filesystem, Edera is able to prevent container escapes. Edera can isolate vulnerabilities in the kernel or drivers to limit their impact, ensuring that only resources within the zone can be affected. 

%% file: discussion.tex
\section{Discussion}
\label{section:discussion}

We discuss addition considerations when using Edera, and opportunities for future work.

\subsection{Zone Flexibility}
\label{section:flexibility}
Edera allows for flexibility through dynamic resource allocation.
Memory and CPU allocations for a zone can be dynamically updated, allowing for more efficient use of these resources and greater flexibility for changing zone requirements.
If a zone starts with 8GB of memory but is running out, the zone can be given additional memory without interfering with the workload.
In most existing systems, allocating more memory requires re-creation of the container.
Further, a Kubernetes container running in a zone can be dynamically moved to another machine without interrupting the workload.
This is possible because the zone is independent of the host machine and so can be moved to other hardware if needed.

\subsection{Usability}
The good performance of Edera will allow organizations to use security mechanisms for all containers, not just those related to security.
For performance reasons, many organizations use OS virtualization for most operations, and a system with stronger isolation (like gVisor or Kata Containers) for certain, security critical applications.
This segmentation of operations leaves most workloads vulnerable to container escapes.
Edera obviates the need for this segmentation, providing strong security for all applications.

Edera achieves this by providing a drop-in replacement for the Kubernetes runtime so that it can be used for isolation without sacrificing usability at all other layers of Kubernetes.
Edera can be applied to a Kubernetes configuration with a 1 line change adding the runtime class to the spec section:

\begin{verbatim}
spec:
  runtimeClassName: edera
\end{verbatim}


%
%

\subsection{Future work}
Opportunities for future work include:
\begin{itemize}
    \item Further optimizations to improve Edera's startup and system call performance.
    \item Using Edera for driver isolation. Isolating GPUs and other drivers further reduces the TCB.
    \item Hardening the Edera hypervisor through the use of hardware enclaves or support for confidential computing.
    For example the startup process could be verified by hardware keys stored in an enclave.
    \item 
    A protocol like expanded Berkeley Packet Filtering (eBPF) could be used in the hypervisor to detect anomalies and prevent exploits running in a zone. Further, monitoring of known compromised kernels could be used for compromise detection.
\end{itemize}

%% file: conclusion.tex
\section{Conclusion}
\label{section:conclusion}

We present and evaluate Edera, an optimized VMM that has performance comparable to popular operating system virtualization technologies while providing strong isolation.
We implement Edera along with a Kubernetes compatible container runtime to demonstrate how it can work within existing infrastructure.
We find that Edera outperforms gVisor in CPU performance, memory performance, and under real-world workloads, while only adding 0.74 seconds of startup time without the use of virtualization extensions. 
It further achieves comparable performance to Docker, Kata Containers, and Firecracker.

%% file: appendix-kata.tex
\appendix
\section{Kata Backend Comparison}
\label{appendix:kata}

\begin{figure*}[ht]
	\centering
	\includegraphics[width=\textwidth]{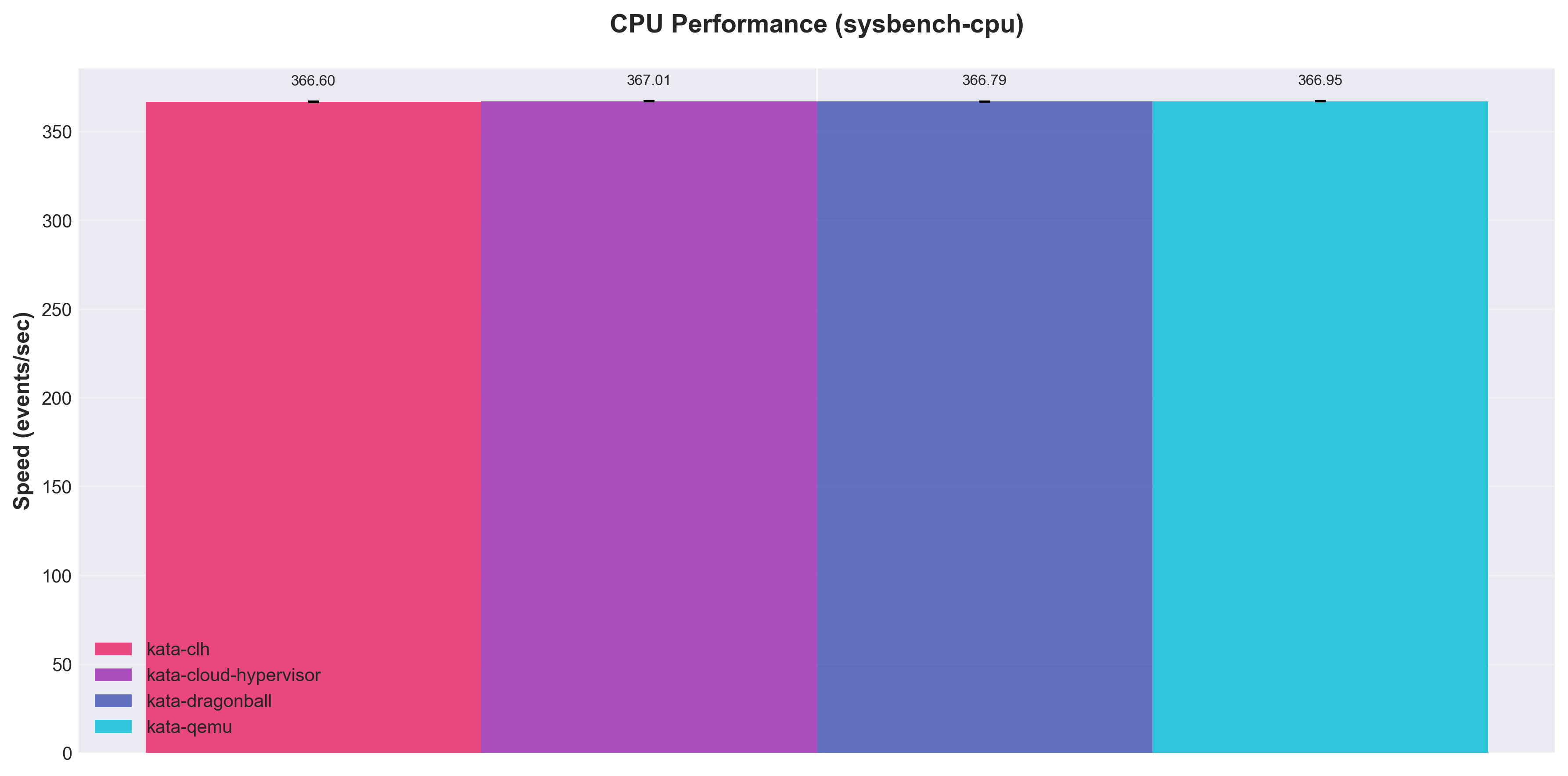}
	\caption{A comparison of the CPU speed in events per second using metrics from sysbench. Higher is better.}
	\label{fig:kata_cpu}
\end{figure*}

\begin{figure*}[ht]
	\centering
	\includegraphics[width=\textwidth]{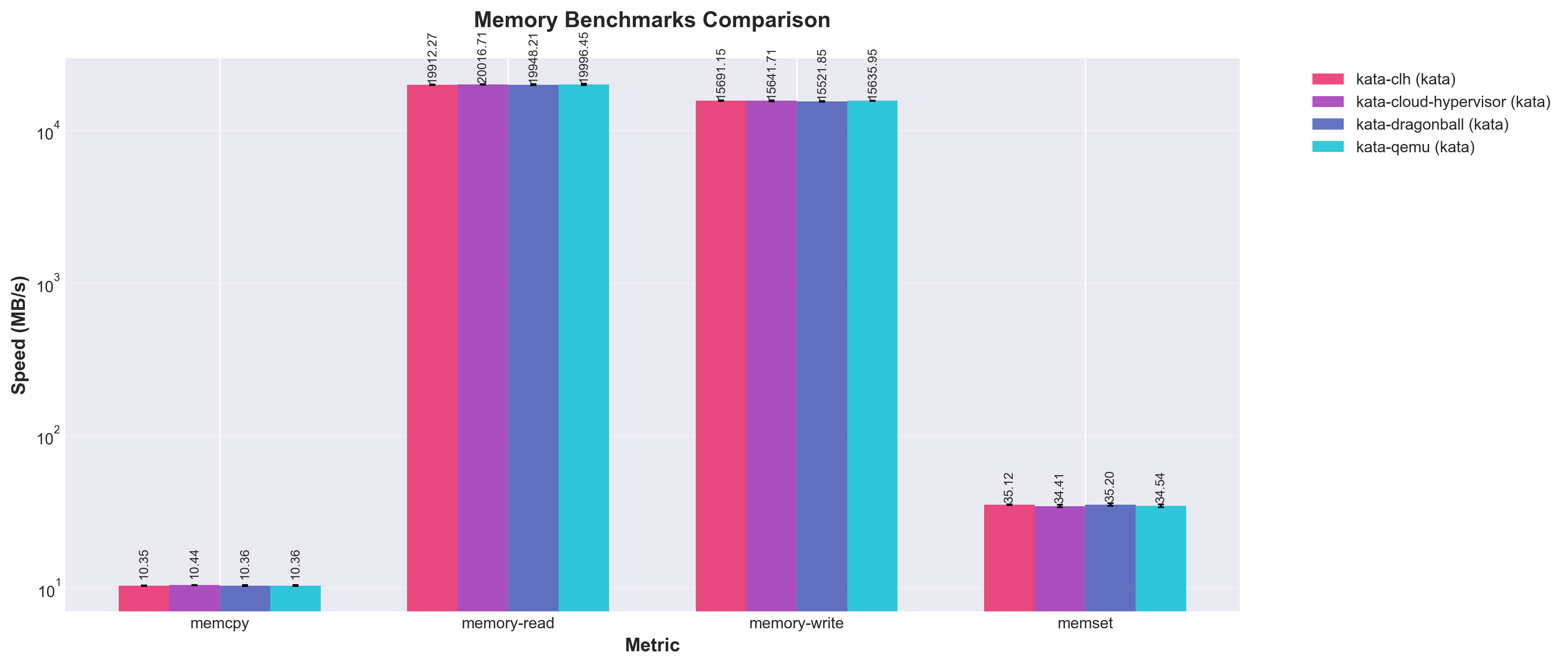}
	\caption{A comparison of the memory speed in GB/s using metrics from both sysbench and perf-bench. Higher is better.}
	\label{fig:kata_memory}
\end{figure*}

\begin{figure*}[t]
	\centering
	\includegraphics[width=\textwidth]{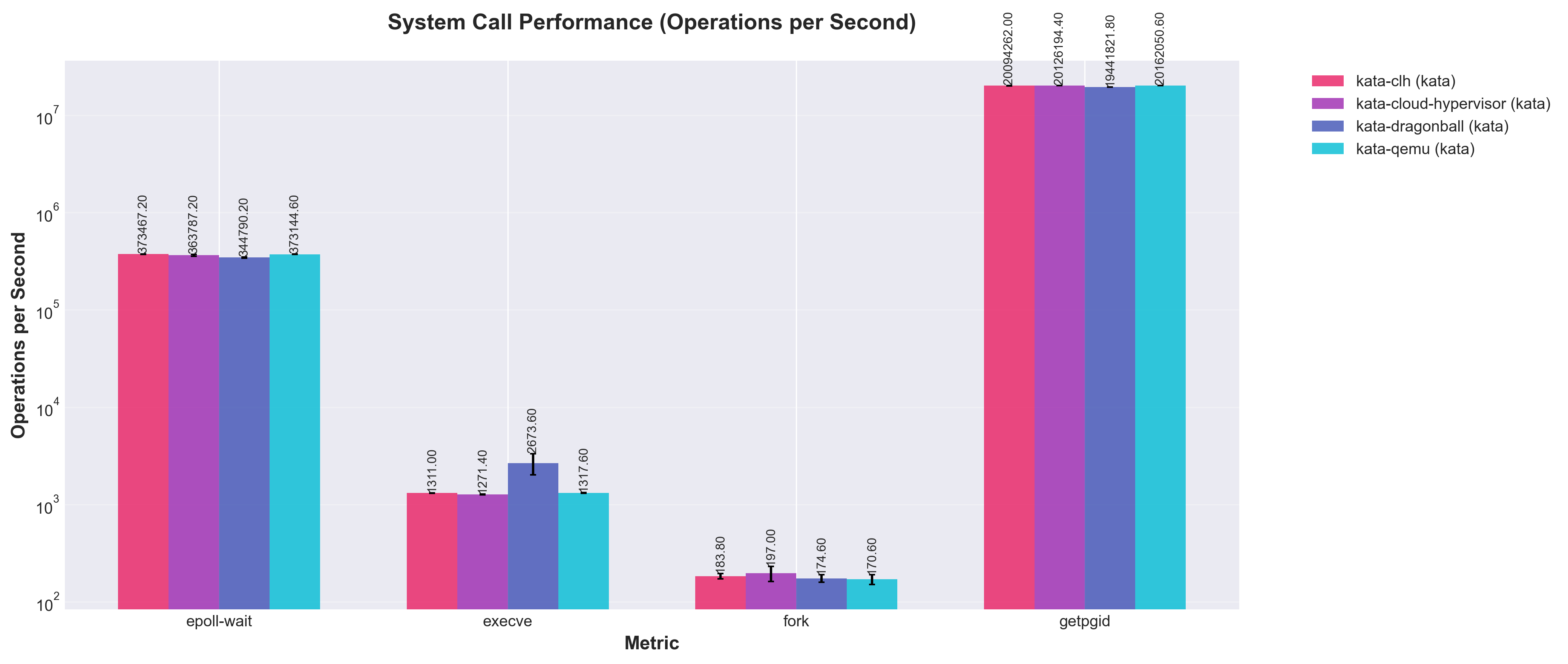}
	\caption{A comparison of the operations per second of system calls shown on a logarithmic scale. Higher is better.}
	\label{fig:kata_syscall}
\end{figure*}

In order to select a Kata backend for the comparisons in this paper, we confirmed which had the best performance in our setup. We tested clh, cloud hypervisor, dragonball, and qemu. We found the results to be very similar, with Dragonball performing slightly better in a few cases. The CPU, memory, and syscall performance from this comparison are shown in \cref{fig:kata_cpu}, \cref{fig:kata_memory}, and \cref{fig:kata_syscall}.

%% file: bibliography.bib
@inproceedings{xen-blanket,
	author = {Williams, Dan and Jamjoom, Hani and Weatherspoon, Hakim},
	title = {The Xen-Blanket: virtualize once, run everywhere},
	year = {2012},
	isbn = {9781450312233},
	publisher = {Association for Computing Machinery},
	address = {New York, NY, USA},
	url = {https://doi.org/10.1145/2168836.2168849},
	doi = {10.1145/2168836.2168849},
	booktitle = {Proceedings of the 7th ACM European Conference on Computer Systems},
	pages = {113–126},
	numpages = {14},
	location = {Bern, Switzerland},
	series = {EuroSys '12}
}

@inproceedings{younge2011analysis,
	title={Analysis of virtualization technologies for high performance computing environments},
	author={Younge, Andrew J and Henschel, Robert and Brown, James T and Von Laszewski, Gregor and Qiu, Judy and Fox, Geoffrey C},
	booktitle={2011 IEEE 4th International Conference on Cloud Computing},
	pages={9--16},
	year={2011},
	organization={IEEE}
}

@article{huber2011evaluating,
	title={Evaluating and Modeling Virtualization Performance Overhead for Cloud Environments.},
	author={Huber, Nikolaus and von Quast, Marcel and Hauck, Michael and Kounev, Samuel},
	journal={Closer},
	volume={11},
	pages={563--573},
	year={2011}
}

@misc{gartner-container-usage,
    author       = {Deborah Galea and Neil Carpenter},
    title        = {Container Security Best Practices: Securing Build to Runtime (and Back)},
    howpublished = {\url{https://orca.security/resources/blog/container-security-best-practices/}},
  note         = {Accessed 2024-10-15},
  year={2024},
}

@MISC{CVE-2022-0847,
    key = {CVE-2022-0847},
  title = {{CVE}- 2022-0847.},
  month=mar # "~10",
  year = {2022},
  url={https://nvd.nist.gov/vuln/detail/CVE-2022-0847},
  urldate={15 October 2024}
}

@MISC{CVE-2022-0492,
    key = {CVE-2022-0492},
  title = {{CVE}- 2022-0492.},
  month=feb # "~6",
  year = {2022},
  url={https://access.redhat.com/security/cve/cve-2022-0492},
  urldate={15 October 2024}
}

@MISC{CVE-2022-0185,
    key = {CVE-2022-0185},
  title = {{CVE}- 2022-0185.},
  month=feb # "~11",
  year = {2022},
  url={https://nvd.nist.gov/vuln/detail/CVE-2022-0185},
  urldate={15 October 2024}
}

@MISC{CVE-2022-23648,
    key = {CVE-2022-23648},
  title = {{CVE}- 2022-23648.},
  month=mar # "~3",
  year = {2022},
  url={https://nvd.nist.gov/vuln/detail/CVE-2022-23648},
  urldate={15 October 2024}
}

@MISC{CVE-2022-0811,
    key = {CVE-2022-0811},
  title = {{CVE}- 2022-0811.},
  month=mar # "~16",
  year = {2022},
  url={https://nvd.nist.gov/vuln/detail/CVE-2022-0811},
  urldate={15 October 2024}
}

@MISC{CVE-2024-0132,
    key = {CVE-2024-0132},
  title = {{CVE}- 2024-0132.},
  month=sep # "~26",
  year = {2022},
  url={https://nvd.nist.gov/vuln/detail/CVE-2024-0132},
  urldate={15 October 2024}
}

@techreport{Denali,
  title={Denali: Lightweight Virtual Machines for Distributed and Networked Applications},
  author={Andrew Whitaker and Marianne Shaw and Steven D. Gribble},
  year={2001},
    institution = {The University of Washington},
  url={https://www.semanticscholar.org/paper/Denali\%3A-Lightweight-Virtual-Machines-for-and-Whitaker-Shaw/11928ecc96f52e153f6a3bf5143260f15f7c4dfd}
}

@article{xen,
author = {Barham, Paul and Dragovic, Boris and Fraser, Keir and Hand, Steven and Harris, Tim and Ho, Alex and Neugebauer, Rolf and Pratt, Ian and Warfield, Andrew},
title = {Xen and the art of virtualization},
year = {2003},
issue_date = {December 2003},
publisher = {Association for Computing Machinery},
address = {New York, NY, USA},
volume = {37},
number = {5},
issn = {0163-5980},
url = {https://doi.org/10.1145/1165389.945462},
doi = {10.1145/1165389.945462},
journal = {SIGOPS Oper. Syst. Rev.},
month = oct,
pages = {164–177},
numpages = {14}
}

@inproceedings{Unikernels:paper,
author = {Madhavapeddy, Anil and Mortier, Richard and Rotsos, Charalampos and Scott, David and Singh, Balraj and Gazagnaire, Thomas and Smith, Steven and Hand, Steven and Crowcroft, Jon},
title = {Unikernels: library operating systems for the cloud},
year = {2013},
isbn = {9781450318709},
publisher = {Association for Computing Machinery},
address = {New York, NY, USA},
url = {https://doi.org/10.1145/2451116.2451167},
doi = {10.1145/2451116.2451167},
pages = {461–472},
numpages = {12},
location = {Houston, Texas, USA},
booktitle = {ASPLOS '13}
}

@inproceedings{lightvm,
author = {Manco, Filipe and Lupu, Costin and Schmidt, Florian and Mendes, Jose and Kuenzer, Simon and Sati, Sumit and Yasukata, Kenichi and Raiciu, Costin and Huici, Felipe},
title = {My VM is Lighter (and Safer) than your Container},
year = {2017},
isbn = {9781450350853},
publisher = {Association for Computing Machinery},
address = {New York, NY, USA},
url = {https://doi.org/10.1145/3132747.3132763},
doi = {10.1145/3132747.3132763},
booktitle = {Proceedings of the 26th Symposium on Operating Systems Principles},
pages = {218–233},
numpages = {16},
location = {Shanghai, China},
series = {SOSP '17}
}

@inproceedings{unikraft,
author = {Kuenzer, Simon and B\u{a}doiu, Vlad-Andrei and Lefeuvre, Hugo and Santhanam, Sharan and Jung, Alexander and Gain, Gaulthier and Soldani, Cyril and Lupu, Costin and Teodorescu, \c{S}tefan and R\u{a}ducanu, Costi and Banu, Cristian and Mathy, Laurent and Deaconescu, R\u{a}zvan and Raiciu, Costin and Huici, Felipe},
title = {Unikraft: fast, specialized unikernels the easy way},
year = {2021},
isbn = {9781450383349},
publisher = {Association for Computing Machinery},
address = {New York, NY, USA},
url = {https://doi.org/10.1145/3447786.3456248},
doi = {10.1145/3447786.3456248},
booktitle = {Proceedings of the Sixteenth European Conference on Computer Systems},
pages = {376–394},
numpages = {19},
location = {Online Event, United Kingdom},
series = {EuroSys '21}
}

@inproceedings{hermitux,
author = {Olivier, Pierre and Chiba, Daniel and Lankes, Stefan and Min, Changwoo and Ravindran, Binoy},
title = {A binary-compatible unikernel},
year = {2019},
isbn = {9781450360203},
publisher = {Association for Computing Machinery},
address = {New York, NY, USA},
url = {https://doi.org/10.1145/3313808.3313817},
doi = {10.1145/3313808.3313817},
booktitle = {Proceedings of the 15th ACM SIGPLAN/SIGOPS International Conference on Virtual Execution Environments},
pages = {59–73},
numpages = {15},
location = {Providence, RI, USA},
series = {VEE 2019}
}

@inproceedings {firecracker,
author = {Alexandru Agache and Marc Brooker and Alexandra Iordache and Anthony Liguori and Rolf Neugebauer and Phil Piwonka and Diana-Maria Popa},
title = {Firecracker: Lightweight Virtualization for Serverless Applications },
booktitle = {17th USENIX Symposium on Networked Systems Design and Implementation (NSDI 20)},
year = {2020},
isbn = {978-1-939133-13-7},
address = {Santa Clara, CA},
pages = {419--434},
url = {https://www.usenix.org/conference/nsdi20/presentation/agache},
publisher = {USENIX Association},
month = feb
}

@inproceedings{pvm,
author = {Huang, Hang and Lai, Jiangshan and Rao, Jia and Lu, Hui and Hou, Wenlong and Su, Hang and Xu, Quan and Zhong, Jiang and Zeng, Jiahao and Wang, Xu and He, Zhengyu and Han, Weidong and Liu, Jiang and Ma, Tao and Wu, Song},
title = {PVM: Efficient Shadow Paging for Deploying Secure Containers in Cloud-native Environment},
year = {2023},
isbn = {9798400702297},
publisher = {Association for Computing Machinery},
address = {New York, NY, USA},
url = {https://doi.org/10.1145/3600006.3613158},
doi = {10.1145/3600006.3613158},
booktitle = {Proceedings of the 29th Symposium on Operating Systems Principles},
pages = {515–530},
numpages = {16},
location = {Koblenz, Germany},
series = {SOSP '23}
}

@ARTICLE{vkernel,
  author={Huang, Hang and Wang, Honglei and Rao, Jia and Wu, Song and Fan, Hao and Yu, Chen and Jin, Hai and Suo, Kun and Pan, Lisong},
  journal={IEEE Transactions on Computers}, 
  title={vKernel: Enhancing Container Isolation via Private Code and Data}, 
  year={2024},
  volume={73},
  number={7},
  pages={1711-1723},
  doi={10.1109/TC.2024.3383988}}

@misc{oxide,
    author       = {Oxide},
    title        = {Oxide Cloud Computer},
    howpublished = {\url{https://oxide.computer/}},
  note         = {Accessed 2024-10-16},
  year={2024},
}

@misc{metalvisor,
    author       = {Mainsail},
    title        = {Metalvisor: Introducing the first TypeZero Hypervisor for Next-Gen Security \& Performance},
    howpublished = {\url{https://www.mainsailindustries.com/metalvisor}},
  note         = {Accessed 2024-10-16},
  year={2024},
}

@misc{constellation,
    author       = {{Constellation} maintainers},
    title        = {Constellation: Always Encrypted Kubernetes},
    howpublished = {\url{https://github.com/edgelesssys/constellation}},
  note         = {Accessed 2024-10-16},
  year={2024},
}

@misc{kubelet,
    author       = {{Kubernetes}},
    title        = {kubelet},
    howpublished = {\url{https://kubernetes.io/docs/reference/command-line-tools-reference/kubelet/}},
  note         = {Accessed 2024-10-17},
  year={2024},
}

@inproceedings {popular-paths,
author = {Yiwen Li and Brendan Dolan-Gavitt and Sam Weber and Justin Cappos},
title = {{Lock-in-Pop}: Securing Privileged Operating System Kernels by Keeping on the Beaten Path},
booktitle = {2017 USENIX Annual Technical Conference (USENIX ATC 17)},
year = {2017},
isbn = {978-1-931971-38-6},
address = {Santa Clara, CA},
pages = {1--13},
url = {https://www.usenix.org/conference/atc17/technical-sessions/presentation/li-yiwen},
publisher = {USENIX Association},
month = jul
}

@misc{xen-channel,
    author       = {{Xen Project} maintainers},
    title        = {Xen PV Channels},
    howpublished = {\url{https://xenbits.xenproject.org/docs/unstable/man/xen-pv-channel.7.html}},
  note         = {Accessed 2024-10-21},
  year={2024},
}

@misc{prometheus,
    author       = {{Prometheus} Authors},
    title        = {Prometheus},
    howpublished = {\url{https://prometheus.io/}},
  note         = {Accessed 2024-10-21},
  year={2024},
}

@inproceedings {solaris-zones,
author = {Daniel Price and Andrew Tucker},
title = {Solaris Zones: Operating System Support for Consolidating Commercial Workloads},
booktitle = {18th Large Installation System Administration Conference (LISA 04)},
year = {2004},
address = { Atlanta, GA},
url = {https://www.usenix.org/conference/lisa-04/solaris-zones-operating-system-support-consolidating-commercial-workloads},
publisher = {USENIX Association},
month = nov
}

@inproceedings {apiary,
author = {Shaya Potter and Jason Nieh},
title = {Apiary: {Easy-to-Use} Desktop Application Fault Containment on Commodity Operating Systems},
booktitle = {2010 USENIX Annual Technical Conference (USENIX ATC 10)},
year = {2010},
url = {https://www.usenix.org/conference/usenix-atc-10/apiary-easy-use-desktop-application-fault-containment-commodity-operating},
publisher = {USENIX Association},
month = jun
}

@misc{namespaces,
    author       = {{Linux} maintainers},
    title        = {namespaces(7) - Linux manual page},
    howpublished = {\url{https://man7.org/linux/man-pages/man7/namespaces.7.html}},
  note         = {Accessed 2024-10-23},
  year={2024},
}

@misc{cgroups,
    author       = {{Linux} maintainers},
    title        = {cgroups(7) - Linux manual page},
    howpublished = {\url{https://man7.org/linux/man-pages/man7/cgroups.7.html}},
  note         = {Accessed 2024-10-23},
  year={2024},
}

@misc{seccomp,
    author       = {{Linux} maintainers},
    title        = {seccomp(2) - Linux manual page},
    howpublished = {\url{https://man7.org/linux/man-pages/man2/seccomp.2.html}},
  note         = {Accessed 2024-10-23},
  year={2024},
}

@misc{capabilities,
    author       = {{Linux} maintainers},
    title        = {capabilities(7) - Linux manual page},
    howpublished = {\url{https://man7.org/linux/man-pages/man7/capabilities.7.html}},
  note         = {Accessed 2024-10-23},
  year={2024},
}

@misc{selinux,
    author       = {{SELinux} maintainers},
    title        = {SELinux Userspace},
    howpublished = {\url{https://github.com/SELinuxProject/selinux}},
  note         = {Accessed 2024-10-16},
  year={2024},
}

@misc{lxc,
    author       = {{LXC} maintainers},
    title        = {What's LXC},
    howpublished = {\url{https://linuxcontainers.org/lxc/introduction/}},
  note         = {Accessed 2024-10-16},
  year={2024},
}

@misc{kubernetes,
    author       = {{Kubernetes} maintainers},
    title        = {Kubernetes},
    howpublished = {\url{https://kubernetes.io/}},
  note         = {Accessed 2024-10-28},
  year={2024},
}

@misc{cri,
    author       = {{Kubernetes} maintainers},
    title        = {Container Runtime Interface (CRI)},
    howpublished = {\url{https://kubernetes.io/docs/concepts/architecture/cri/}},
  note         = {Accessed 2024-10-28},
  year={2024},
}

@misc{mirage,
    author       = {{MirageOS} maintainers},
    title        = {A programming framework for building type-safe, modular systems},
    howpublished = {\url{https://mirage.io/}},
  note         = {Accessed 2024-10-28},
  year={2024},
}

@misc{kata,
    author       = {{Kata Containers} maintainers},
    title        = {The speed of containers, the security of VMs},
    howpublished = {\url{https://katacontainers.io/}},
  note         = {Accessed 2024-10-28},
  year={2024},
}

@misc{infranetes,
    author       = {{Infranetes} maintainers},
    title        = {Infranetes},
    howpublished = {\url{https://github.com/apporbit/infranetes}},
  note         = {Accessed 2024-10-28},
  year={2016},
}

@misc{rkt,
    author       = {{rkt} maintainers},
    title        = {Running rkt with KVM stage1},
    howpublished = {\url{https://github.com/rkt/rkt/blob/master/Documentation/running-kvm-stage1.md}},
  note         = {Accessed 2024-10-28},
  year={2019},
}

@misc{virtlet,
    author       = {{virtlet} maintainers},
    title        = {virtlet},
    howpublished = {\url{https://github.com/Mirantis/virtlet}},
  note         = {Accessed 2024-10-28},
  year={2019},
}

@misc{runx,
    author       = {{RunX} maintainers},
    title        = {RunX: next generation secured containers},
    howpublished = {\url{https://xcp-ng.org/blog/2021/09/14/runx-next-generation-secured-containers/}},
  note         = {Accessed 2024-10-28},
  year={2021},
}

@misc{gvisor,
    author       = {{gVisor} maintainers},
    title        = {The Container Security Platform},
    howpublished = {\url{https://gvisor.dev/}},
  note         = {Accessed 2024-10-28},
  year={2024},
}

@inproceedings{cooperative-paravirtualization,
author = {Liu, Yuxuan and Xu, Tianqiang and Mi, Zeyu and Hua, Zhichao and Zang, Binyu and Chen, Haibo},
title = {CPS: A Cooperative Para-virtualized Scheduling Framework for Manycore Machines},
year = {2024},
isbn = {9798400703942},
publisher = {Association for Computing Machinery},
address = {New York, NY, USA},
url = {https://doi.org/10.1145/3623278.3624762},
doi = {10.1145/3623278.3624762},
booktitle = {Proceedings of the 28th ACM International Conference on Architectural Support for Programming Languages and Operating Systems, Volume 4},
pages = {43–56},
numpages = {14},
location = {Vancouver, BC, Canada},
series = {ASPLOS '23}
}

@inproceedings{paravirt-numa,
  title={When extended para-virtualization (XPV) meets NUMA},
  author={Bui, Bao and Mvondo, Djob and Teabe, Boris and Jiokeng, Kevin and Wapet, Lavoisier and Tchana, Alain and Thomas, Ga{\"e}l and Hagimont, Daniel and Muller, Gilles and DePalma, Noel},
  booktitle={Proceedings of the Fourteenth EuroSys Conference 2019},
  pages={1--15},
  year={2019}
}

@inproceedings{xen-64,
    author = {Jun Nakajima and Asit Mallick and Ian Pratt and Keir Fraser},
    title = {X86-64 XenLinux: Architecture, Implementation, and
Optimizations} ,
    booktitle = {2006 Linux Symposium},
    year = 2006
}

@misc{gvisor-gemini-attack,
    author       = {Roni Carta},
    title        = {We hacked Google’s A.I Gemini and leaked its source code (at least some part)},
    howpublished = {\url{https://www.landh.tech/blog/20250327-we-hacked-gemini-source-code/}},
  note         = {Accessed 2025-04-23},
  year={2025},
}

@misc{kvm,
    author       = {{Linux} Developers},
    title        = {Kernel Virtual Machine},
    howpublished = {\url{https://linux-kvm.org/page/Main_Page}},
  note         = {Accessed 2025-04-23},
  year={2025},
}

@inproceedings {operation-forwarding,
author = {Jietao Xiao and Nanzi Yang and Wenbo Shen and Jinku Li and Xin Guo and Zhiqiang Dong and Fei Xie and Jianfeng Ma},
title = {Attacks are Forwarded: Breaking the Isolation of {MicroVM-based} Containers Through Operation Forwarding},
booktitle = {32nd USENIX Security Symposium (USENIX Security 23)},
year = {2023},
isbn = {978-1-939133-37-3},
address = {Anaheim, CA},
pages = {7517--7534},
url = {https://www.usenix.org/conference/usenixsecurity23/presentation/xiao-jietao},
publisher = {USENIX Association},
month = aug
}

@misc{confidential-containers,
    author       = {{Confidential Containers} maintainers},
    title        = {Confidential Containers},
    howpublished = {\url{https://github.com/confidential-containers}},
  note         = {Accessed 2025-04-23},
  year={2025},
}

@misc{cncf-project-metrics,
    author       = {{Cloud Native Computing Foundation}},
    title        = {Project Metrics},
    howpublished = {\url{https://www.cncf.io/project-metrics/}},
  note         = {Accessed 2025-04-23},
  year={2025},
}

@misc{xen-aerospace,
    author       = {{Dorner Works}},
    title        = { Enable interoperability and software security for embedded products with DornerWorks Xen Project Hypervisor},
    howpublished = {\url{https://www.dornerworks.com/solutions/embedded-virtualization/virtuosity/}},
  note         = {Accessed 2025-05-01},
  year={2025},
}

@misc{xen-automotive,
    author       = {{epam}},
    title        = {Automotive Virtualization by Xen},
    howpublished = {\url{https://www.epam.com/industries/industrial/automotive/xen-virtualization}},
  note         = {Accessed 2025-05-01},
  year={2025},
}

@misc{xenproject-embedded,
    author       = {{Xen Project}},
    title        = {Embedded \& Automotive},
    howpublished = {\url{https://xenproject.org/projects/embedded-and-automotive/}},
  note         = {Accessed 2025-05-01},
  year={2025},
}

@inproceedings {memory-ballooning,
	author = {Jui-Hao Chiang and Han-Lin Li and Tzi-cker Chiueh},
	title = {Working Set-based Physical Memory Ballooning},
	booktitle = {10th International Conference on Autonomic Computing (ICAC 13)},
	year = {2013},
	isbn = {978-1-931971-02-7},
	address = {San Jose, CA},
	pages = {95--99},
	url = {https://www.usenix.org/conference/icac13/technical-sessions/presentation/chiang},
	publisher = {USENIX Association},
	month = jun
}

@misc{xen-numa,
    author       = {{Xen Project}},
    title        = {Xen on NUMA Machines},
    howpublished = {\url{https://wiki.xenproject.org/wiki/Xen_on_NUMA_Machines}},
  note         = {Accessed 2025-05-01},
  year={2025},
}

@misc{kata-hypervisors,
    author       = {{Kata Containers Maintainers}},
    title        = {Hypervisors},
    howpublished = {\url{https://github.com/kata-containers/kata-containers/blob/main/docs/hypervisors.md}},
  note         = {Accessed 2025-05-01},
  year={2025},
}
